\documentclass[superscriptaddress,twocolumn,amsmath,amssymb,aps,pre,floatfix]{revtex4}

\usepackage{comment}
\usepackage[english]{babel}
\usepackage[utf8]{inputenc}
\usepackage[T1]{fontenc}
\usepackage{booktabs}
\usepackage{xr}
\usepackage{float}
\usepackage{natbib}
\usepackage{units}
\usepackage{graphicx}
\usepackage{animate}
\usepackage{dcolumn}
\usepackage{bm}
\usepackage{hyperref}
\hypersetup{colorlinks=true, linktoc=all, linkcolor=blue, linktocpage, citecolor=blue}

\newcommand{\deta}{\Delta\eta}

\begin{document}
\title{Opinion inertia and coarsening in the Persistent Voter model} 

\author{Luís Carlos F. Latoski}
\email{luis.latoski@ufrgs.br}
\affiliation{Instituto de Física, Universidade Federal do Rio Grande do Sul, CEP 91501-970, Porto Alegre - RS, Brazil} 
\author{W. G. Dantas}
\email{wgdantas@id.uff.br}
\affiliation{Departamento de Ciências Exatas, EEIMVR, Universidade Federal Fluminense, CEP 27255-125, Volta Redonda - RJ, Brazil}
\author{Jeferson J. Arenzon}
\email{arenzon@if.ufrgs.br}
\affiliation{Instituto de Física, Universidade Federal do Rio Grande do Sul, CEP 91501-970, Porto Alegre - RS, Brazil} 
\affiliation{Instituto Nacional de Ciência e Tecnologia - Sistemas Complexos, Rio de Janeiro - RJ, Brazil}
\date{\today}

\begin{abstract}
We consider the Persistent Voter model  (PVM), a variant of the Voter model (VM) that includes transient, dynamically-induced zealots.
Due to peer reinforcement, the internal confidence $\eta_i$ of a normal voter increases by steps of size $\deta$ and once it gets above a given threshold, it  becomes a zealot.
Then, its opinion remains frozen until enough interactions with the opposite opinion occur and its confidence is reset.
No longer a zealot, the regular voter may change opinion once again.
This opinion inertia mechanism, albeit simplified, is responsible for an effective surface tension and the PVM has a crossover from a fluctuation-driven dynamics, as in the VM, to a curvature-driven one, as in the Ising Model at low temperature (IM0). 
The average time $\tau$ to attain consensus is non-monotonic on $\deta$ and has a minimum at $\deta_{\min}$.
In this paper we clarify the mechanisms that accelerate the system towards consensus close  to $\deta_{\min}$.
Close to the crossover at $\deta_{\min}$, the intermediate region around the domains where the regular voters accumulate (the active region, AR) is large and the surface tension, albeit small, is still enough to keep the shape and reduce the fragmentation of the domains.
The large size of the AR in the region of  $\deta_{\min}$ has two important effects that accelerates the dynamics.
First, it dislodges the zealots in the bulk of the domains and second, it maximally suppresses the slowly-evolving stripes that normally form in Ising-like models. 
This suggests the importance of understanding the role of the AR, where the change of opinion is facilitated, and the interplay between regular voters and zealots when attempting to disrupt polarized states.
\end{abstract}

\maketitle

\section{Introduction}
\label{sec.intro}

Consensus~\cite{Baronchelli18} is an emergent property resulting from multiple interactions in a collection of agents  where observation and imitation lead to social learning.
It is a  characteristic of a society that, through the repeated exchange between the individuals, has arrived to a major agreement on important issues.
Social-influence models like those considered here are mostly based on positive influence, with individuals adapting their opinions to those prevalent among their neighbors.
Attaining consensus thus seems a natural consequence of the collective action of agents.
However, many factors may deter or delay such process, sometimes leading to polarization, with the population divided into two or more antagonistic positions.
In this case, local consensus is possible as there is a tendency to minimize in-group dissent while global consensus is prevented because out-group differences intensify.
Many models have studied these processes with either discrete or continuous opinions~\cite{CaFoLo09,Baronchelli18,Redner19}, attempting to both understand the underlying mechanisms and reproduce the experimental observations~\cite{MaPl13,GrSuRaMiEg14}.
The Voter model (VM)~\cite{ClSu73,HoLi75,Redner01,Redner19} has agents choosing from a discrete set of opinions (we here consider the binary case) and imitating, in each step, one of their neighbors irrespective of how different their opinions are.
Its dynamics is thus driven by interfacial fluctuations~\cite{DoChChHi01}.
This is different from some spin magnetic models where local consensus (domains of connected, parallel spins) grows because of the surface tension and the minimization of the interfaces~\cite{Bray94}. 
An example is the zero-temperature dynamics of the Ising model from a random initial state~\cite{Bray94} (referred here as IM0).
Before entering the regime where the movement of the interfaces between opposite spins is driven by curvature, these 2d systems are initially often brought close to the random percolation critical point~\cite{ArBrCuSi07,SiArBrCu07,BlCuPiTa17,AzAlOlAr22}.
The geometric details of the first stable percolating cluster determine the asymptotic state of the system for zero temperature dynamics~\cite{SpKrRe01a,BaKrRe09,OlKrRe12}, whether fully-magnetized (consensus) or divided into parallel stripes of opposite spins, a segregated and polarized state where the horizontal and vertical symmetry is broken~\cite{SpKrRe01a,SpKrRe01b,BaKrRe09,CaBaLo09,OlKrRe12,VoRe12}.
Some of these results have been experimentally verified in the ordering kinetics of liquid crystals~\cite{SiArDiBrCuMaAlPi08,AlTa21,Almeida23}, whose dynamics with a non-conserved order parameter is believed to be in the corresponding Ising dynamical universality class~\cite{Bray94,AlTa21}.
Because of the emergent surface tension in the model considered here, the phenomenology of the curvature-driven systems will be relevant in the following sections.

Although in the original VM the agents have no memory beyond the previous step, we are interested in  extensions where the opinions have inertia and their switching probability evolves in time. 
Refs.~\cite{StTeSc08,StTeSc08b,PeKhRo20}, among others, considered inertia by changing the flipping probability as a function of the time elapsed since the last flip.
It is also possible to add a latent period, after each flip, when the flipping probability remains low before eventually getting back to the usual voter behavior~\cite{LaSaBl09,WaLiWaZhWa14}.
Another form of inertia includes different levels of confidence on a given opinion. 
Upon interaction with the opposite view, instead of a sudden change of opinion as in the VM, the agents may decrease its confidence~\cite{VoRe12,VeVa18} or first become  undecided, bearing an intermediate, neutral position~\cite{VaKrRe03,BaPiSe15,SvSw15}.
In many of these examples, although the microscopic dynamics was slowed down, there is a macroscopic acceleration and it is possible to reach the consensus faster than in the original VM.

One interesting limit is when inertia dominates and the opinions of some agents remain frozen, the so-called zealot strategy~\cite{Mobilia03,MoPeRe07,GaJa07,MaGa13,CoCa16}.
Recently, the non-markovian Persistent Voter Model  (PVM) was introduced and studied~\cite{LaDaAr22}.
Because of the successive reinforcement, long enough persistent opinions may turn a regular voter into a zealot, blocking its flipping capability. 
However, this is a transient state: upon contact with a different opinion, its behavior is reset to a normal voter.
This zealot-like behavior of highly confident agents is somewhat akin to close mindedness, where opinions have an inertia and need multiple interactions with different agents to be overcomed.
Open mindedness, or easy flipping, occurs in the normal voter model where a single contact with a different opinion is enough for an agent to change state.
The time persistence induced by the internal confidence makes more difficult to conform with the fluctuating opinions around the agent, helping to reduce polarization.
Indeed, the presence of these self-induced frozen agents delays the opinion change, what helps to avoid polarization  and may accelerate the dynamics toward consensus. 
This is at odds with the behavior of non-transient zealots that, instead, may hinder consensus~\cite{MoGe05,MoPeRe07}.

In this work we further explore the PVM properties, uncovering quantitative similarities with the dynamics of the IM0, despite being a non-Markovian, detailed balance-violating, non-Hamiltonian model.
In particular, some of the questions we attempt to answer are the following.
How does the inertia to change opinion affect the time to attain a global consensus?
When inertia induces an effective surface tension, how similar is the dynamics to the IM0?
How is the crossover from the fluctuation to the curvature-driven regime?
If the exit time has a minimum, as in other models, which is the responsible mechanism?
Opposite zealots tend to segregate in the bulk of domains, with normal voters occupying the interstitial regions.
How important is the width of these regions, and the internal dynamics, for the evolution of consensus?
The paper is organized as follows.
Section~\ref{sec.model} defines the PVM while the simulation results are presented in Sect.~\ref{sec.results}, first for specially prepared ordered initial states and then for the more general case, with random initial configurations. 
These results are then discussed in Sect.~\ref{sec.conclusions}.

\section{The Persistent Voter Model}
\label{sec.model}

In the VM, the opinion of the $i$-th agent ($i=1, \ldots, N$) is represented by a binary variable $s_i=\pm 1$ and can be shared with its neighbors.
The Persistent Voter model (PVM)~\cite{LaDaAr22} extends it to include an internal, positive and continuous variable, $\eta_i\geq 0$, associated with the individual confidence on its own opinion. 
When this confidence gets above a given threshold, $\eta_i\geq\phi$ (set, from now on, to $\phi=1$), the opinion becomes frozen and the agent behaves as a zealot.
Although the opinion of a zealot is not affected by the neighboring opinions, its associated confidence may be.
The dynamics is as follows.
A focal agent $i$ and one of its nearest neighbors $j$ are chosen.
If $i$ is not a zealot, $\eta_i<1$, it follows the usual imitation dynamics of the VM and its opinion $s_i$ aligns with $s_j$, if $s_i\neq s_j$.
Despite of what happens with $s_i$, both $\eta_i$ and $\eta_j$ evolve.
If $s_i$ flips, it is because of the influence of $j$.
On the other hand, if $s_i$ is already equal to $s_j$, that reinforces the confidence of $j$.
In either case, $\eta_j$ increases, $\eta_j   \longrightarrow \eta_j + \deta$ with $\deta>0$.
For the focal agent $i$, if both opinions were the same, the mutual reinforcement is positive, otherwise, this single confrontation with a different opinion is enough to reset its confidence (see Ref.~\cite{LaDaAr22} for a more general version of the model):
\begin{align}
\eta_i  \longrightarrow
\left\{
\begin{array}{ll}
        \eta_i + \deta, & \text{if } s_i=s_j\\
                        & \\
         0,              & \text{if } s_i \neq s_j.
\end{array}
\right.
\end{align}
As usual, one Monte Carlo step (MCS) consists in repeating this procedure $N$ times. 

The only  parameter is the confidence increment $\Delta\eta$ that controls how fast the threshold to become a zealot is approached.
For small $\deta$, the formation of zealots is delayed and the agent behaves as a regular voter for a longer time.
On the other hand, when $\deta$ is large enough, zealots easily nucleate in the center of domains, away from agents with an opposite opinion, inducing an effective curvature-driven dynamics at the interfaces in spite of its roughness.
As a consequence, on regular 2d lattices there is an intermediate time regime in which the model behavior resembles~\cite{LaDaAr22} the zero-temperature kinetic Ising Model (IM0)~\cite{Bray94}.
It is the interplay between the mechanisms that underlie both the Ising and the Voter models that is responsible for the interesting properties of the PVM discussed in the next sections.
In some limiting cases, the PVM becomes very similar to the marginal model of Ref.~\cite{VoRe12} and the $M=2$ case of Ref.~\cite{VeVa18}.

\section{Results}
\label{sec.results}

\begin{figure}
\includegraphics[width=.241\columnwidth]{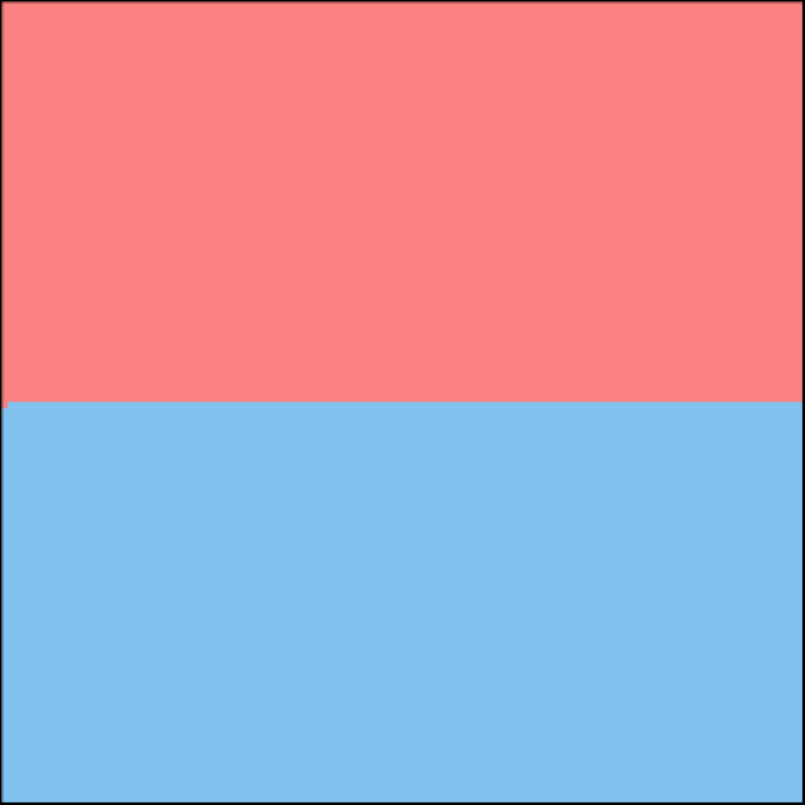}
\includegraphics[width=.241\columnwidth]{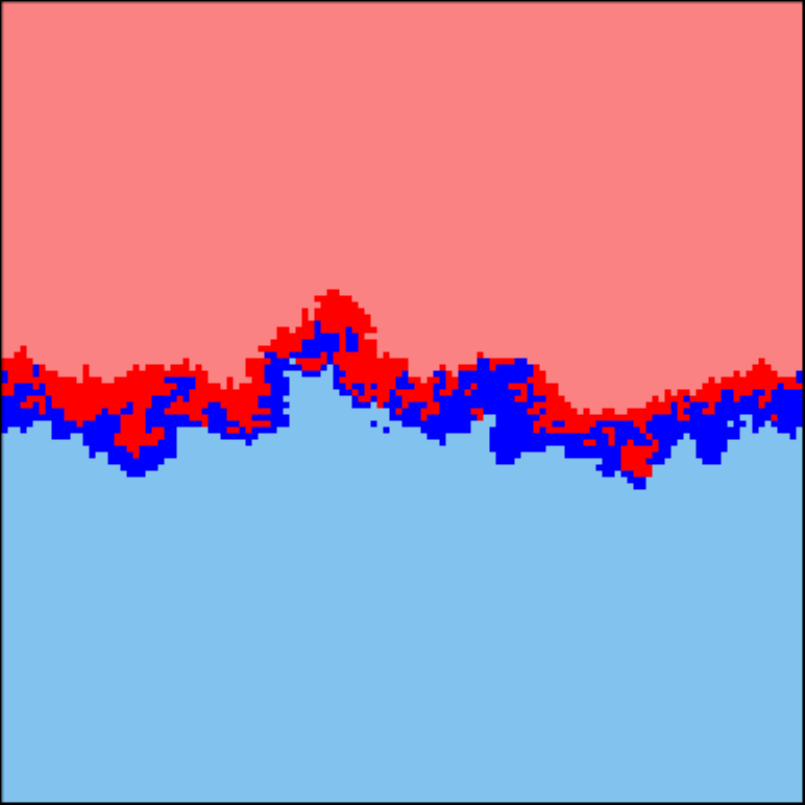}
\includegraphics[width=.241\columnwidth]{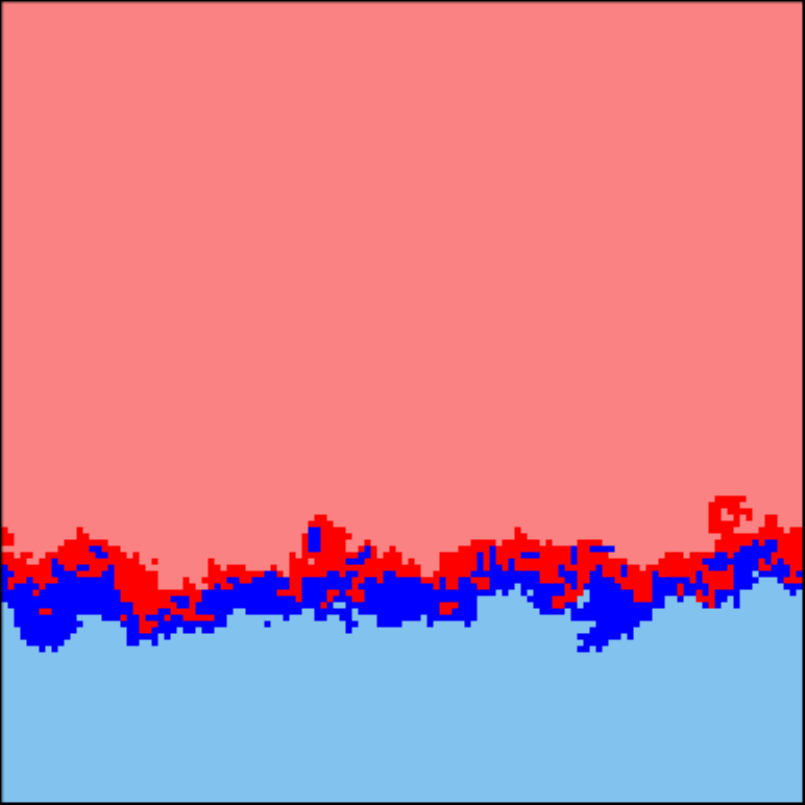}
\includegraphics[width=.241\columnwidth]{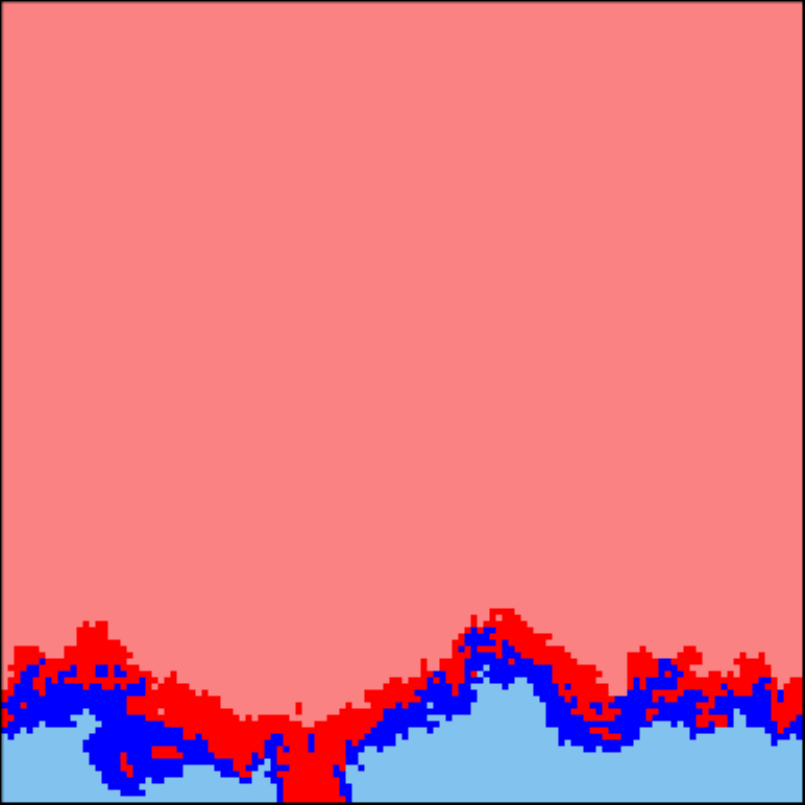}

\vspace{1mm}
\includegraphics[width=.241\columnwidth]{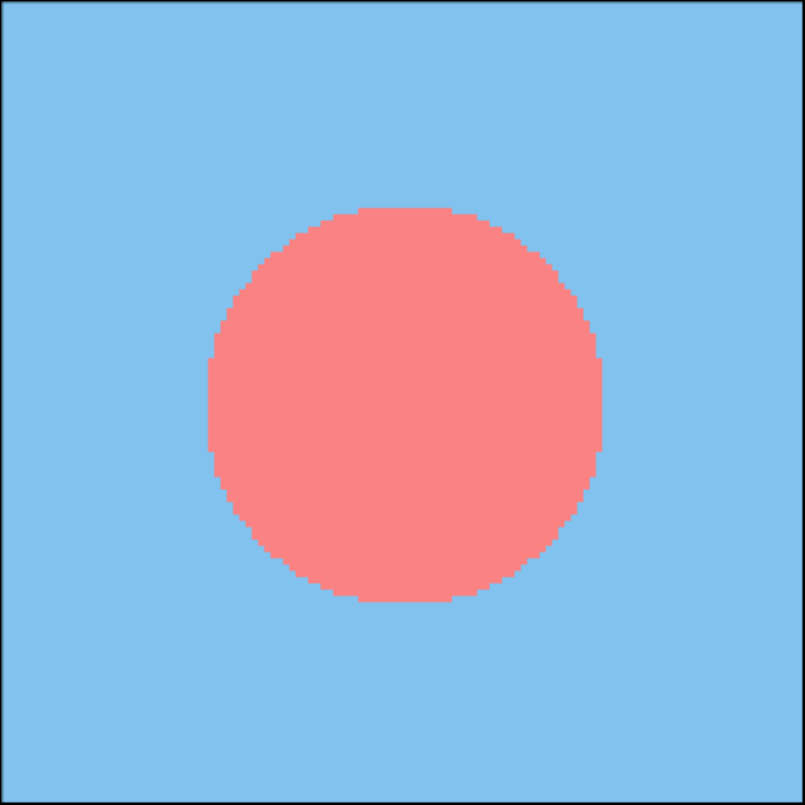}
\includegraphics[width=.241\columnwidth]{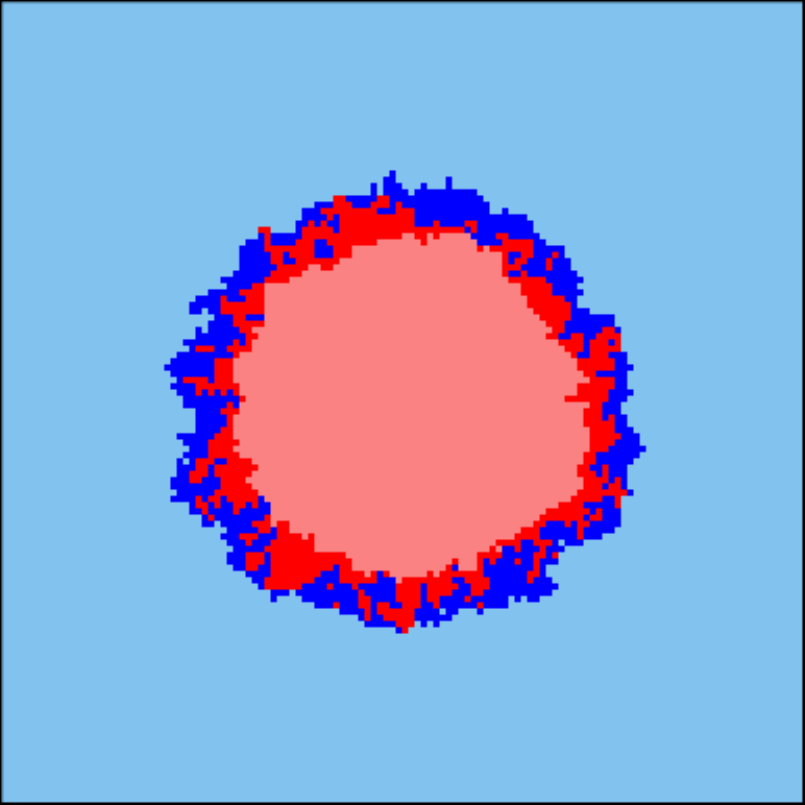}
\includegraphics[width=.241\columnwidth]{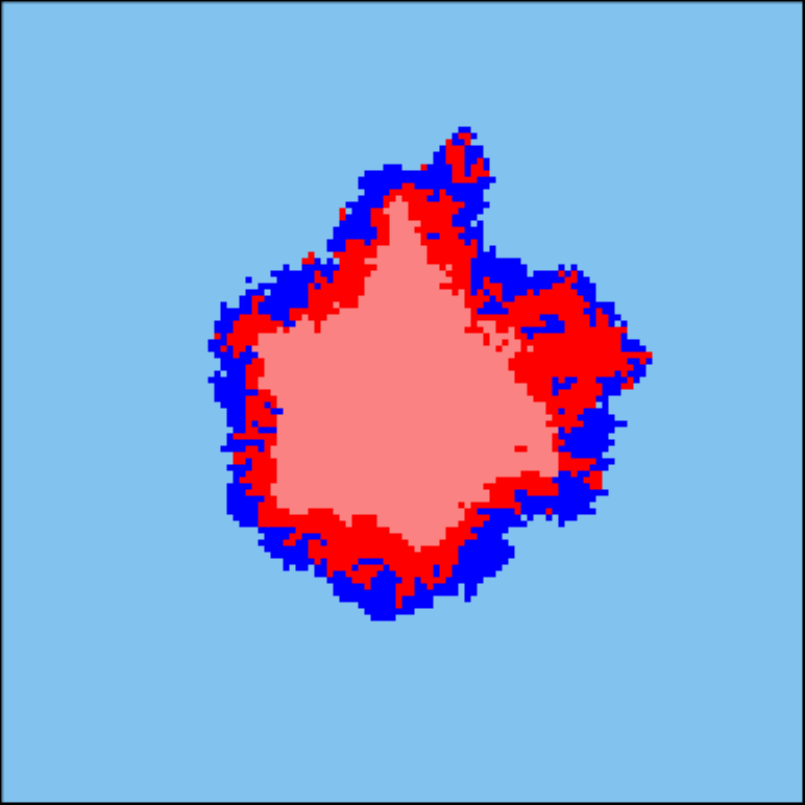}
\includegraphics[width=.241\columnwidth]{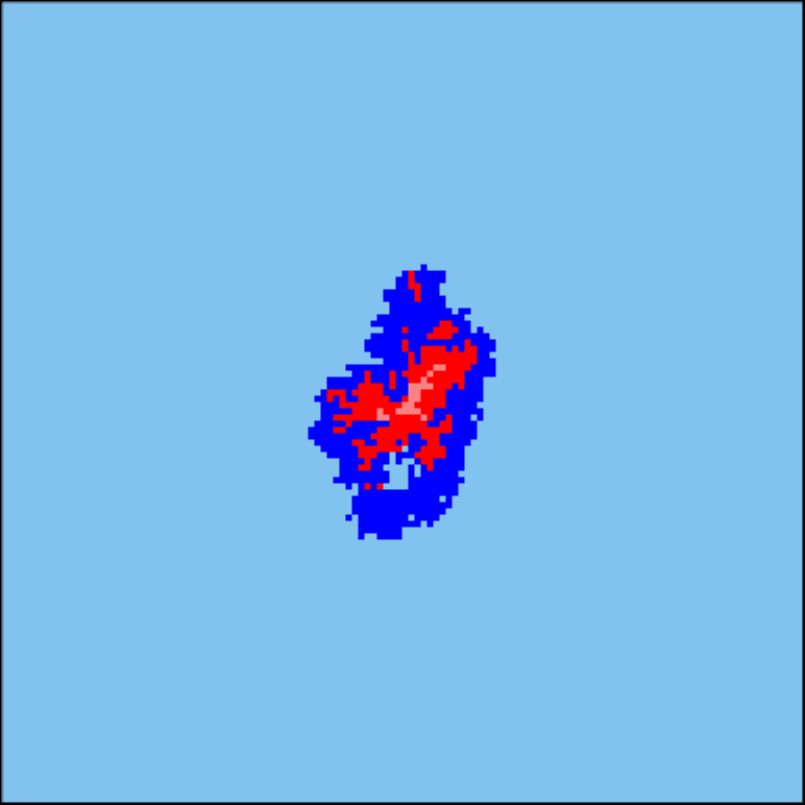}

\vspace{1mm}
\includegraphics[width=.241\columnwidth]{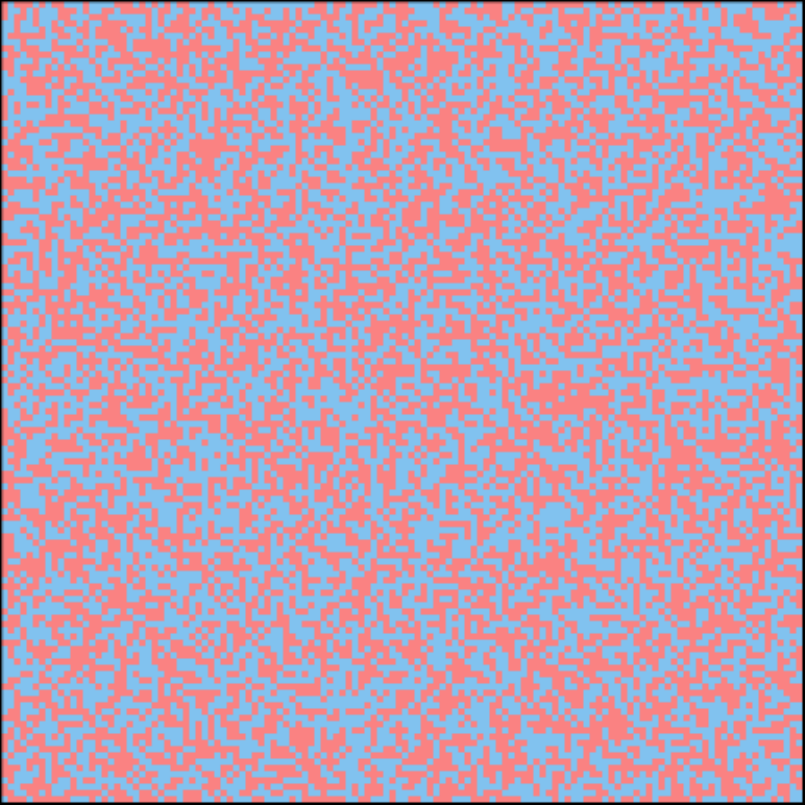}
\includegraphics[width=.241\columnwidth]{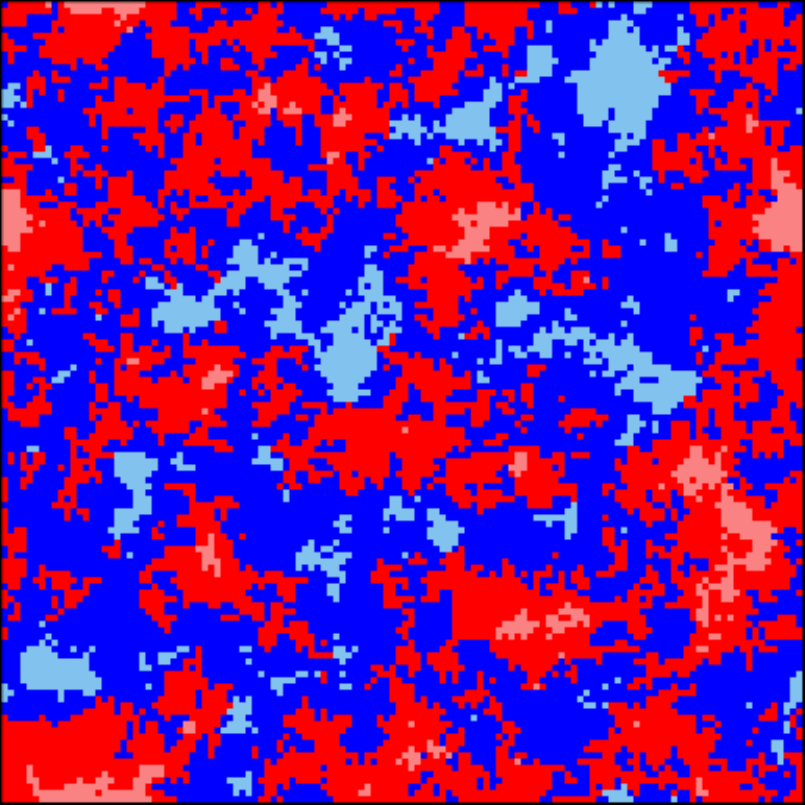}
\includegraphics[width=.241\columnwidth]{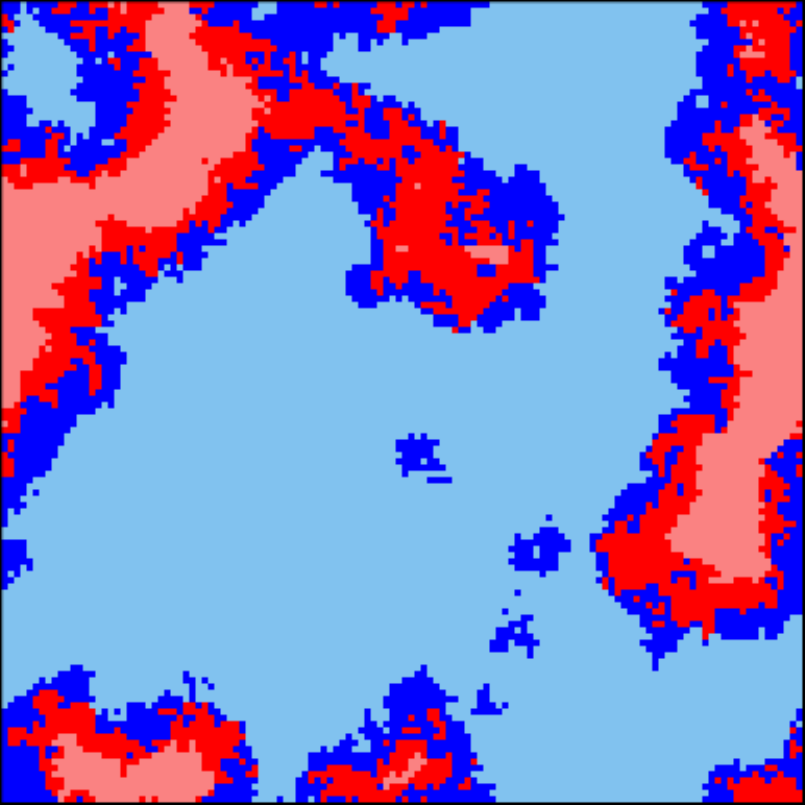}
\includegraphics[width=.241\columnwidth]{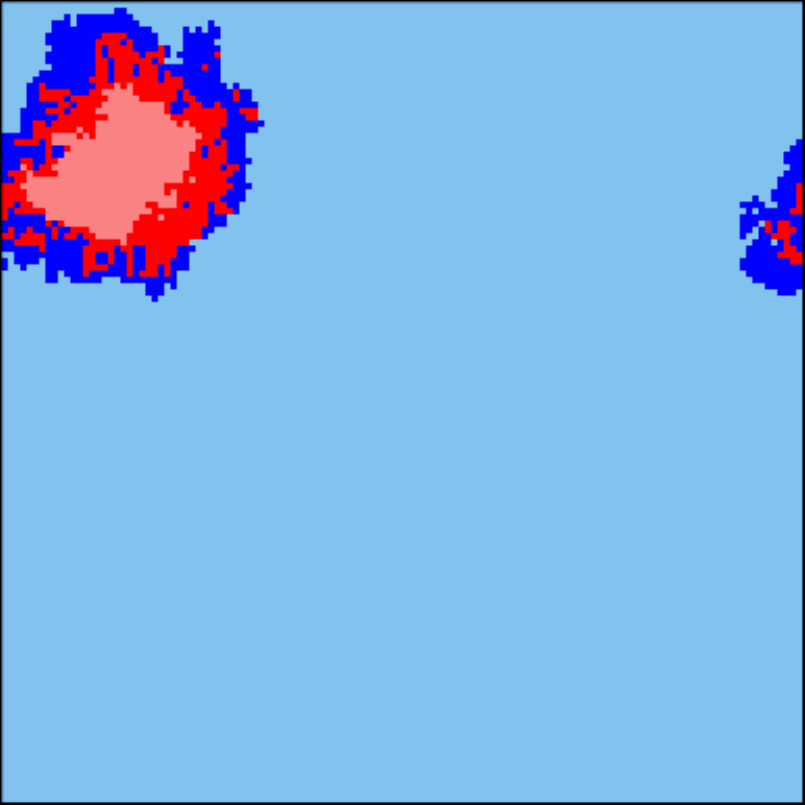}

\caption{Snapshots for the PVM with $\deta=10^{-2}$ and the different initial states discussed in the text. Red and blue indicate the two opposite opinions, with regular voters (zealots) having the darker (lighter) colors. In all cases, zealots segregate and the normal voters occupy the intermediate active region (AR). (Top) Flat initial interface with opposite opinions distributed in equal-sized rectangular stripes.  Once the AR forms, it diffuses through the lattice and eventually collides with the border. After that, consensus is easily attained.
    (Middle) Circular initial interface of radius $R_0=32$ sites separating the group of same-opinion agents inside the circle from those outside. Both the AR and the central region with zealots shrink until they both disappear. 
    (Bottom) Random initial state with each opinion chosen with the same probability. Domains of different sizes form and evolve, resembling the IM0.
    }
    \label{fig.snapshots}
\end{figure}

Non-random initial states with particular geometries help to better understand specific features related to the initial segregation and its further evolution. 
In Sects.~\ref{sec.ordered.flat} and \ref{sec.ordered.circ} we consider two regular initial states, with either a flat or a circular interface, respectively, separating the regions with different opinions on a 2d square lattice of finite size $L$.
Then, in Sect.~\ref{sec.random}, instead of a single pair of tailored domains, we study a random, fully disordered initial configuration where the emergent coarsening dynamics may evolve toward the consensus through configurations that resemble the above circular or flat interfaces.
In all cases, there is a broad range of $\deta$ that induces surface tension in the PVM, even without an energy cost associated with the interface.
These initial states and the associated curvature driven dynamics are exemplified in the sequences shown in the several rows of Fig.~\ref{fig.snapshots}.
Unless explicitly noticed, averages are over sets with $10^3$ or $10^4$ samples.

\subsection{Ordered initial states: flat interface}
\label{sec.ordered.flat}

\begin{figure}[htb]
\includegraphics[width=\columnwidth]{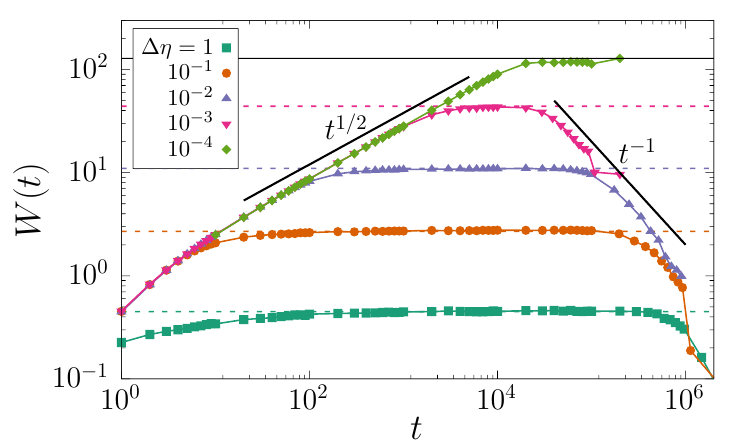}
\caption{Starting from a flat initial interface, the average width $W$ of the AR presents three very distinct regimes. 
For values of $\deta$ not too large, the early behavior becomes diffusive as the reset mechanism unfreezes the system while developing the AR. 
In the second regime, the AR is stationary (horizontal dashed lines). 
The smaller $\deta$ is, the longer the start of the plateau is delayed and the larger $W$ is. 
Interestingly, the time to attain consensus also decreases until $W\sim L$ (in this case, $L=128$, shown as a horizontal solid line), when the VM is recovered and $\tau$ grows again.
Lastly, the third regime is when the system departs from the plateau.
Due to finite-size effects, the AR collapses after touching the border and consensus is soon attained (see the first row in Fig.~\ref{fig.snapshots}).}
\label{fig.wall1}
\end{figure}

A possible initial state with a fully polarized configuration consists of two parallel stripes of opposite opinions, where all agents start as zealots, i.e., $\eta_i=1$, $\forall i$.
Regular voters are created and accumulate in the intermediate, active region (AR) that separates the zealots, shown in Fig.~\ref{fig.snapshots} with darker colors.
In this case, since the interest is in how the initial flat interface between them evolves, we use open boundary conditions.
A similar flat interface would prevent the IM0 from evolving, with both stripes lasting forever.
Analogously, in the PVM with large $\deta$, the system remains mostly pinned because of the large fraction of zealots.
In this case, regular voters turn into zealots almost immediately and the AR becomes very thin and, in some cases, fragmented (e.g., for $\deta=1$ in Fig.~\ref{fig.wall1}, the width is smaller than a single site, $W<1$).
Although small groups of zealots may coexist in the interior of the AR, the average size of the latter may be well approximated by $W\simeq L-z(t)/L$, as shown in Fig.~\ref{fig.wall1}, where $z(t)$ is the average number of zealots at time $t$.

\begin{figure}[htb]
\includegraphics[width=\columnwidth]{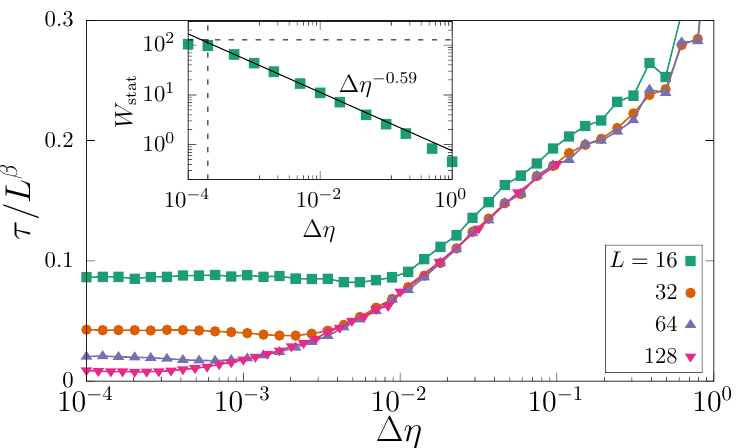}
\caption{The average time $\tau$ to attain consensus, rescaled by $L^{\beta}$, as a function of $\deta$ for a flat initial interface.
There is a minimum of $\tau$ separating the VM-like behavior for $\deta<\deta_{\rm min}$ from the curvature-driven regime for $\deta>\deta_{\rm min}$.
As $L$ increases, $\deta_{\rm min}$ shifts to smaller values as $\deta_{\rm min}\sim L^{-1.5}$.  
Except for the smallest size, we obtain an excellent collapse for $\deta>\deta_{\rm min}$ with $\beta\simeq 3.2$.
At $\deta_{\rm min}$, on the other hand, $\tau$ seems to scale as $L^{2.06}$ (not shown), indicating very different underlying mechanisms.
Inset:  the power-law behavior of the width of the AR in the stationary regime, $W_{\rm stat}\sim (\deta)^{-0.59}$. The minimum average time $\tau$ to attain consensus (vertical dashed line) occurs close to the point where the width of the AR is approaching $L$ (horizontal dashed line).}
\label{fig.wall2}
\end{figure}

We are interested in those values of $\deta$ for which the AR is not so thin and the emergent surface tension induces novel behavior.
There are many different interfaces in this system, some cross the lattice while others are closed (e.g., those of the small domains inside the AR). 
Their behavior depend on the competition between the zealots formation and the confidence resetting, and set the three distinct time regimes of the AR  shown in Fig.~\ref{fig.wall1}: 1) initial diffusive growth,  2) stationary width and, eventually, 3) finite-size instability and collapse (leading to consensus).
The early development of the AR as $W\sim t^{1/2}$ is solely related to the diffusive dynamics of the interfaces and is mostly independent of $\deta$.
The AR increases because the zealots at the border interact with agents with the opposite opinion, becoming regular voters after their confidence is reset.
A second interaction is necessary for the opinion to change, but it depends on the next-return time of one of the interfaces while it performs a confined random-walk inside the AR.
The confidence keeps increasing because of the continued interactions with like-minded neighbours and if that second contact takes too long, the voter may become a zealot again, decreasing the AR.
In this way, the inward movement of the zealots interfaces that closes the AR competes with the outward collisions with the opposite opinion that increase it.
When the two timescales are similar, the AR attains its stationary width $W_{\rm stat}$, whose power-law increase,  $W_{\rm stat} \sim (\deta)^{-0.59}$,
is shown in the inset of Fig.~\ref{fig.wall2} for $L=128$.
By decreasing the value of $\deta$, the formation of zealots is delayed.
Consequently, the departure from the diffusive regime occurs later and the AR becomes larger.
For $\deta$ small enough, the AR occupies the whole lattice, $W_{\rm stat}\sim L$, and one recovers the VM (e.g., $\deta\simeq 10^{-4}$ in Fig.~\ref{fig.wall1}).

Once fully developed, in the stationary regime, the width of the AR remains stable for a considerable amount of time while  it diffuses through the lattice, Fig.~\ref{fig.snapshots} (top row).
The departure of the plateau is a percolation event, triggered when one of the opinions breaks and destabilizes the opposite zealots stripe, spanning the whole lattice in both directions.
This percolating cluster may form multiple times, but once stable, its opinion invades the system very fast, as seen by the $t^{-1}$ envelope in Fig.~\ref{fig.wall1}.
As $\deta$ decreases, the AR gets wider and closer to the border, decreasing the consensus time.
The minimum time is attained soon before the AR occupies the whole system.
At $\deta_{\rm min}$, the intermediate regime (the plateau) disappears and the time to consensus is dominated by the slower diffusive process that builds the AR, implying in $\tau_{\rm min}\sim L^2$.
Once $W_{\rm stat}\sim L$, zealots are very rare and one recovers the VM, the plateau reappears and lasts a long time.
In Fig.~\ref{fig.wall2} we show the average time $\tau$ to attain consensus, noticing indeed the existence of a minimum value, $\tau_{\rm min}\equiv\tau(\deta_{\rm min})$.
For larger values of $L$, $\tau_{\rm min}$ increases while $\deta_{\rm min}$ becomes smaller.
A good collapse around the minimum is obtained rescaling $\tau$ and $\deta$ with  $\tau_{\rm min} \sim L^{2.06}$ and $\deta_{\rm min} \sim L^{-1.5}$, respectively (not shown).
For values $\deta\lesssim \deta_{\rm min}$, the behavior corresponds to the VM and is mostly independent of $\deta$.
On the other hand, for $\deta>\deta_{\rm min}$, except for the smaller lattice ($L=16$), all other sizes collapse very well when $\tau$ is rescaled by $L^{\beta}$, with $\beta\simeq 3.2$. 
The different exponents in these regions indicate that the underlying mechanisms ruling the approach to consensus are different, as will become clear after the discussion in the following sections.

\begin{figure}
\includegraphics[width=\columnwidth]{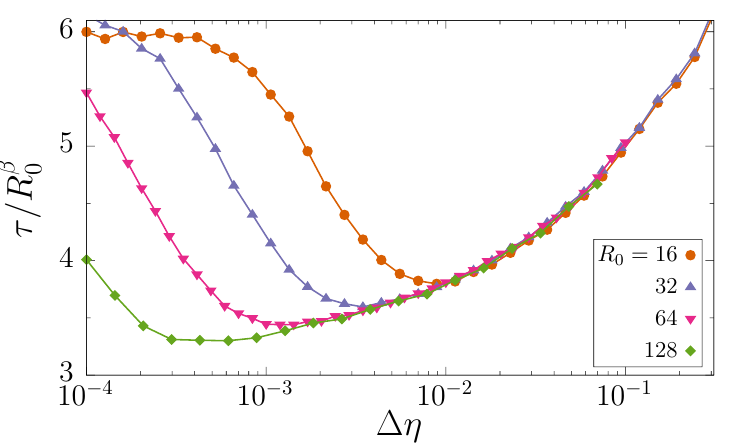}
\caption{Rescaled average consensus time $\tau$ of a single droplet embedded on a large lattice as a function of $\deta$ for different initial radius $R_0$. The location of the minimum deviates toward smaller values, $\deta_{\min}\sim R_0^{-1.5}$, as in the previous case. The best collapse in the region of the minimum was obtained with $\beta\simeq 1.95$, while the value used in the figure, that gives an excellent collapse to the right of $\deta_{\min}$ is $\beta\simeq 2.01$.}
    \label{fig.timeradius}
\end{figure}

\subsection{Ordered initial states: circular interface}
\label{sec.ordered.circ}

Further insight can be gained from another controlled, non-random initial configuration: a single droplet of radius $R_0$ embedded in a sea of the opposite opinion~\cite{DoChChHi01}. 
The effective surface tension constrains the cluster fragmentation to the AR region around the circle,  favoring a more rounded cluster.
This is illustrated in the middle row of Fig.~\ref{fig.snapshots} for $\deta=10^{-2}$.
In comparison with the flat interface discussed in the previous section, the mean curvature accelerates the approach to consensus.
Fig.~\ref{fig.timeradius} shows the average consensus  time $\tau$ as a function of $\deta$ for several values of $R_0$, the relevant linear scale for this case.
In spite of the different geometry, there is again a non-monotonous behavior and $\tau$ presents a minimum, i.e., there is an optimal amount of zealots that, by not being willing to flip, helps to accelerate the dynamics.
For $\deta\gg\deta_{\rm min}$, the dynamics is slower due to the excess of zealots.
Reducing $\deta$, the approach to consensus is faster because the AR increases, dislodging the zealots inside the circle.
The minimum occurs when the size of the AR almost coincides with the whole drop, the residual surface tension is barely enough to reduce the dispersal and the corresponding zealots inside the circle disappears.
Decreasing $\deta$ even further, $\deta\ll\deta_{\rm min}$, the droplet grows, becomes fragmented and invades the whole system.
The large number of normal voters in this case turns the dynamics slower, similarly to the VM.
Moreover, an excellent collapse is observed in Fig.~\ref{fig.timeradius}, $\tau\sim R_0^{\beta}$  with $\beta\simeq 2.01$ and $\deta_{\rm min}\sim R_0^{-1.5}$.
However, $\tau_{\rm min}$ seems to grow with a slightly different exponent, $\tau_{\rm min}\sim R_0^{1.95}$.
This difference is probably due to finite-size effects.

Finally, in Ref.~\cite{LaDaAr22} it was shown for the PVM  that the evolution of the disk area, $A(t)$, in analogy with Ising-like models, may decrease linearly with time when the surface tension becomes important, $A(t)=A(0)-\lambda t$, where $\lambda$ is a function of $\deta$.
It went unnoticed, however, that in the region close to $\deta_{\rm min}$ there seems to be a close correspondence with the Ising model in the same conditions, i.e., $\lambda\simeq 1$ (since the PVM needs two steps to flip a zealot, it is twice as slow as the 2d Ising model at $T=0$).
However, when the initial IM0 state is random, the phenomenology is richer than the evolution of a single, isolated domain.
For example, the system approaches a critical percolating state early in the dynamics~\cite{ArBrCuSi07,SiArBrCu07,AzAlOlAr22} and whose characteristics determine whether consensus is approached fast (as in the drop case) or slower (as in the flat interfaces).
Thus, it is important to understand what is the effect of a distribution of different domain sizes and shapes on the average time to consensus.
This is discussed in the next section.

\subsection{Random initial states}
\label{sec.random}

\begin{figure}[htb]
\includegraphics[width=\columnwidth]{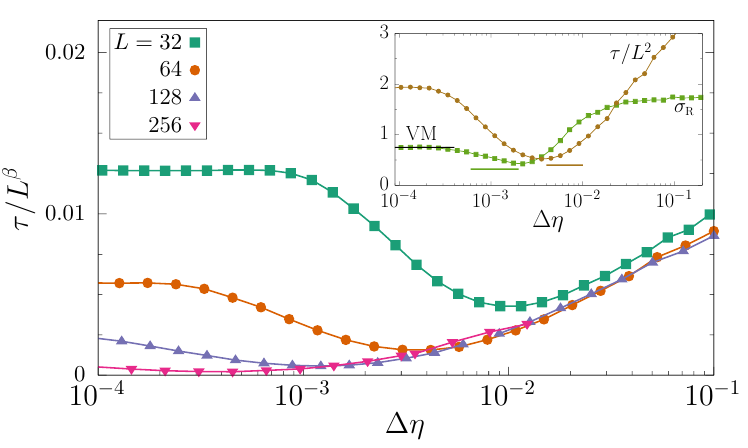}
\includegraphics[width=0.975\columnwidth]{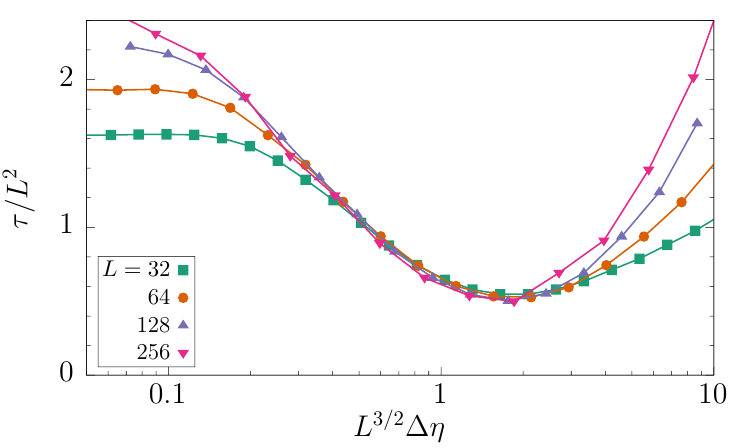}
\caption{Rescaled average consensus time $\tau$ as a  function of $\deta$ for different system sizes and a random initial state.
The two panels show different scalings that collapse either the region around $\deta_{\min}$ (bottom) and $\deta>\deta_{\min}$ (top).
In the top panel, a good collapse is obtained with $\beta\simeq 3.4$.
In analogy with the previous cases, the collapse in the bottom panel was obtained with $\tau_{\min}\sim L^2$ and $\deta_{\min}\sim L^{-3/2}$. 
Inset: Relative standard deviation $\sigma_{\scriptscriptstyle\rm R}$ and $\tau$ as a function of $\deta$ for $L=64$ showing non-monotonic behavior. 
At $\deta_{\min}$ where $\tau$ is minimum, the width of the distribution is also close to its minimum value.  
At that points, these quantities are quite similar to the IM0 when considering, for the latter, only those initial states that converge fast to consensus, without building intermediate stripes. These values are indicated by small horizontal lines (for $\tau$ it was doubled, see text). For very small values of $\deta$, the $\sigma_{\scriptscriptstyle\rm R}$ also coincides with the VM, what is indicated by a black, horizontal strainght line.
}
    \label{fig.timerandoms}
\end{figure}

To avoid correlations between the agents in the initial state, we randomly choose $s_i=\pm 1$ with equal probability while the confidence variables are all set to $\eta_i = 0$.
An example with $\deta=10^{-2}$ is shown in Fig.~\ref{fig.snapshots}, bottom row.
Again, the effective surface tension forms domains whose evolution is similar to the IM0.
As in the previous sections, the average exit time $\tau$ is a non-monotonic function of $\deta$ and the results for several lattice sizes $L$ appear collapsed in Fig.~\ref{fig.timerandoms}.
Around $\deta_{\min}$, a good collapse (bottom panel) is obtained with $\deta_{\min}\sim L^{-3/2}$ and $\tau_{\min}\sim L^2$ (although the minima are wide and small variations in these exponents lead to similarly good collapses).
On the other hand, beyond this point, $\deta>\deta_{\min}$, the size dependence of $\tau$ is given by the larger exponent $\beta\simeq 3.4$ (top panel), indicating that the processes leading to consensus are distinct in those regions. 
This value is similar to those found in the literature for different but related models: 3.4 in the Majority Voter~\cite{ChRe05}, 3.5 in the IM0~\cite{Lipowski99,SpKrRe01a,GoPl18} (in this case, diagonal stripes are the relevant ones), 3.6 in some language models~\cite{CaEgSa06} and the Confident Voter model~\cite{VoRe12}, etc.
A simple argument, considering stripes as a collection of independent 1d random-walkers has been put forward in Ref.~\cite{ChRe05}, obtaining $L^3$. 
Despite close, the values found in the above models are consistently larger.
Velásquez-Rojas et al~\cite{VeVa18} improved on that result by including correlations along the stripe, finding additional corrections and an effective exponent larger than 3.
A complete explanation for these values seems to still be missing.

The relative standard deviation $\sigma_{\scriptscriptstyle\rm R} \equiv \sigma/\tau$, where $\sigma$ is the standard deviation, gives information on the width of the distribution of consensus time, compared with the average.
It also has a non-monotonic behavior as $\deta$ increases, shown in the inset of Fig.~\ref{fig.timerandoms}.
As it was hinted in the previous section, there is a quantitative similarity between the PVM and the IM0 close to the minimum, with each timestep of the latter corresponding to two steps of the former.
For the PVM with random initial conditions, both $\tau$ and $\sigma_{\scriptscriptstyle\rm R}$ have a minimum that seem to be indeed very close to the related values for the IM0 (indicated by straight horizontal lines in the inset).
However, the comparison holds when only a subset of the initial conditions for the IM0 are considered, as we discuss below.

\begin{figure}[htb]
\includegraphics[width=\columnwidth]{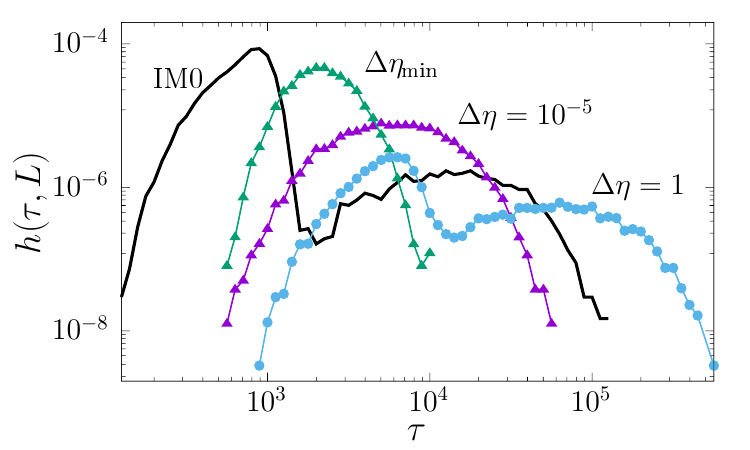}
\caption{Distributions $h(\tau)$ of consensus time $\tau$ for different values of $\deta$, $L=64$ and a random initial state. For comparison, the IM0 distribution is shown (black, solid curve). Notice that the value of $\tau$ corresponding to the first peak is very close to the peak of the $\deta_{\min}$ distribution (see previous section). The results for $\deta=10^{-5}$, with a single, broad peak, is close to the original VM.
}
\label{fig.taudistributions}
\end{figure}

The distribution $h(\tau)$ of the exit time for different values of $\deta$ is shown in Fig.~\ref{fig.taudistributions} for $L=64$. 
For comparison, we also included (solid black line) the two-peaked distribution for the IM0. 
For large $\deta$ (e.g., $\deta=1$), the overall shape of $h(\tau)$ also has two peaks~\cite{LaDaAr22}, but displaced to larger times because the small width of the AR and the expressive number of zealots slows down the dynamics.
In both cases, because of the emergent surface tension and the induced curvature-driven dynamics, a random initial state is attracted to the  percolation critical point early in the dynamics~\cite{ArBrCuSi07}.
Consequently, the geometry of the first stable percolating cluster dictates the timescale for attaining consensus.
The larger peak, at smaller times, corresponds to those initial states that formed, in the beginning of the dynamics, a percolating cluster that wrapped the lattice in both directions.
Instead, those initial states that form  stripes~\cite{Lipowski99,SpKrRe01a}, and whose evolution toward consensus is much slower, contribute to the second peak.
However, in the IM0, initial states that form diagonal stripes are included, but those parallel to the lattice directions (see Ref.~\cite{GoPl18} and references therein for a more general discussion) are not since they freeze in a polarized state and do not evolve to  full consensus.
In the small $\deta$ limit (e.g., $\deta=10^{-5}$), because of the fluctuation-driven dynamics of the AR, stripes are unstable and have no influence whatsoever on the exit time.
The behavior is then closer to the VM, with a single, broad peak.
Between these two extremes, there is, however, an intermediate value, $\deta_{\min}$, where at the same time stripes are unstable but there is enough surface tension to accelerate the dynamics and attaining the consensus time is faster.
At this value of $\deta$, the distribution is close to its smallest width and the peak corresponds to a value that is roughly twice as large as the average $\tau$ for the IM0, as shown in the inset of Fig.~\ref{fig.timerandoms}.
As discussed in the previous section, this is because many flips in the PVM take two steps since, for zealots, the confidence must first be reset. 
As can be observed in Fig.~\ref{fig.taudistributions}, for $\deta_{\min}$, both the location of the peak and the width of the distribution are small, in agreement with Fig.~\ref{fig.timerandoms}.

\section{Conclusions}
\label{sec.conclusions}

We explored the Persistent Voter Model (PVM), an opinion model introduced in Ref.~\cite{LaDaAr22}, analyzing the similarities between the way it approaches consensus and how the zero-temperature Ising model evolves toward a fully magnetized state. 
In the PVM, the confidence $\eta_i$ is an internal attribute of the agent's opinion $s_i$, resulting from the previous interactions with its neighbors.
Above a given threshold for $\eta_i$, the zealot behavior is induced and the opinion of that agent remains frozen.
Such a state is transient and persists while there are no interactions with a different opinion, thus lowering its confidence.
This is very different from bounded-confidence models~\cite{Stone61,DeNeAmWe00,HeKr02}, where the confidence is relative to the opinion being imitated and from which it must not differ too much for a change to occur. 
Although turning a zealot into a normal voter takes at least two steps, the inverse requires $1/\deta$ steps or more.
The parameter $\deta$, in this sense, also sets the degree of asymmetry in the process of changing opinion in the PVM.

Our main results concern the non-monotonic exit time as a function of the reinforcement parameter $\deta$ and the existence of an optimal value for which the approach to consensus is the fastest. 
Consensus time is minimized when the transient polarized states (that, in the IM0, are associated with stripes) are destabilized by the AR around each cluster.
At the same time, the effective surface tension is small to produce a large but confined AR, but enough to dislodge those zealots hidden in the middle of domains, thus accelerating the dynamics.
In other words, around  $\deta>\deta_{\min}$, the noise inherent to the AR that destabilizes the stripes is large but not enough to get rid of a residual surface tension. 
By eliminating the stripes and keeping the curvature-driven dynamics, the path to consensus is faster.
Percolation phenomena  at the early steps of the dynamics have been shown to be essential in framing both the temporal evolution and the asymptotic state of curvature-driven systems like the Ising model~\cite{SpKrRe01a,SpKrRe01b,ArBrCuSi07,SiArBrCu07,BaKrRe09,OlKrRe12,BlCuPiTa17,AzAlOlAr22}
As mentioned in the Introduction, these properties are important also for the PVM and have been experimentally observed in different systems~\cite{SiArDiBrCuMaAlPi08,AlTa21,Almeida23}.
Here we have shown that they are also relevant in a broader context, setting the timescales for an opinion-exchanging population to attain consensus while destabilizing intermediate polarized states.

The effective surface tension generated by the zealots in the PVM is similar but not equivalent to a low, finite temperature in the Ising model.
Indeed, although a small temperature may be useful to  eventually escape from the local minima of flat interfaces, it never allows the system to be fully magnetized.
In the PVM, close to  $\deta_{\min}$, the surface tension is small and the AR acts as an effective noise that increases the roughness of the interfaces.
However, despite large it is not enough to eliminate the curvature-driven decrease of domains toward the absorbing state that is the consensus.

Finally, the simplicity and the interesting behavior of this model may be used to test many important questions.
For example, regarding its geometrical properties, the roughness of the different interfaces along the path to consensus is worth investigating.
Moreover, an important but often neglected question is the correlation between spatial segregation and polarization~\cite{FeFlTo17}, in particular, when zealots are present.  
The bulk segregation easily increases the confidence of agents deep inside the domains, and apparently fosters polarization.
Nonetheless, for specific values of the parameters we observe an accelerated path to consensus.
When does the dynamics induce distancing or reduce spatial sorting of opposite opinions?
Is the level of generated segregation somehow related to the coarsening process of attaining or preventing consensus?
Does segregation between different groups in society foster opinion polarization?
The PVM seems to present a clear pattern of segregation involving polarized opinions and seems a promising model to tackle those questions.

\begin{acknowledgments}
We thank Renan A.L. Almeida for interesting discussions and comments on the manuscript. Work partially supported by the Brazilian agencies FAPERGS, FAPERJ, Conselho Nacional de Desenvolvimento Cientí\-fi\-co e Tecnológico (CNPq), and CAPES. 
\end{acknowledgments}


\end{document}